\documentclass[%
 aapm,
 mph,%
 amsmath,amssymb,
 reprint,%
]{revtex4-2}
\usepackage{mathbbol}
\usepackage{graphicx,color}
\usepackage{amsmath,amssymb,bm,amsthm,amstext}
\usepackage{mathtools}
\usepackage{simplewick}
\usepackage{braket}
\usepackage{qcircuit}
\usepackage{array}

\usepackage{soul}
\usepackage{xfrac}
\usepackage{tikz}
\usepackage{circuitikz}
\usepackage{epstopdf} 
\usepackage{times}
\usepackage{dsfont}
\usepackage[normalem]{ulem}

\usepackage{tabularray}

\usepackage{bm}
\usepackage{mathrsfs}

\DeclareMathAlphabet{\mathtensor}{OT1}{cmss}{sbc}{n}
\DeclareMathAlphabet{\bmathtensor}{OT1}{cmss}{bx}{n}

\renewcommand{\H}{\mathcal{H}}

\newcommand{\e}{\text{e}}

\usepackage[utf8]{inputenc}
\usepackage{graphicx}
\usepackage{dcolumn}
\usepackage{bm}
\usepackage{amsmath}

\usepackage{multirow}

\begin{document}

\preprint{AAPM/123-QED}

\title{Quantum Control of Spin Qubits Using Nanomagnets}

\author{Mohamad Niknam$^\dagger$}
\affiliation{Department of Chemistry and Biochemistry, University of California Los Angeles, 607 Charles E. Young Drive East, Los Angeles, CA 90095-1059, USA}
\altaffiliation[Also at: ]{Center for Quantum Science and Engineering, UCLA}
\author{Md. Fahim F. Chowdhury$^\dagger$}%
\author{Md Mahadi Rajib}
\author{Walid Al Misba}
\affiliation{Department of Mechanical and Nuclear Engineering, College of Engineering, Virginia Commonwealth University,
Richmond, VA 23284-3068, USA}
\author{Robert N. Schwartz}%
\author{Kang L. Wang}%
\affiliation{Device Research Laboratory, Department of Electrical and Computer Engineering, UCLA, 420 Westwood Plaza, Engineering IV,
Los Angeles, CA 90095, USA}
\author{Jayasimha Atulasimha*}
\affiliation{Department of Mechanical and Nuclear Engineering, College of Engineering Virginia Commonwealth University, Richmond, VA 23284-3068, USA}
\altaffiliation[Also at: ]{Department of Electrical and Computer Engineering}
\email{jatulasimha@vcu.edu}
\author{Louis-S. Bouchard*}
\affiliation{Department of Chemistry and Biochemistry, University of California Los Angeles, 607 Charles E. Young Drive East, Los Angeles, CA 90095-1059, USA}
\altaffiliation[Also at: ]{Center for Quantum Science and Engineering, UCLA}
\email{lsbouchard@ucla.edu}

\date{\today}

\begin{abstract}
Single-qubit gates are essential components of a universal quantum computer. Without selective addressing of individual qubits, scalable implementation of quantum algorithms is not possible. When the qubits are discrete points or regions on a lattice, the selective addressing of magnetic spin qubits at the nanoscale remains a challenge due to the difficulty of localizing and confining a classical divergence-free field to a small volume of space. Herein we propose a new technique for addressing spin qubits using voltage-control of nanoscale magnetism, exemplified by the use of voltage control of magnetic anisotropy (VCMA).  We show that by tuning the frequency of the nanomagnet's electric field drive to the Larmor frequency of the spins confined to a nanoscale volume, and by modulating the phase of the drive, single-qubit quantum gates with fidelities approaching those for fault-tolerant quantum computing can be implemented. Such single-qubit gate operations have the advantage of remarkable energy efficiency, requiring only tens of femto-Joules per gate operation, and lossless, purely magnetic field control (no E-field over the target volume).  Their physical realization is also straightforward using existing foundry manufacturing techniques.

\end{abstract}

\keywords{Spin qubit, mesoscopic qubit, quantum control, nanomagnet}
\maketitle

\section{Introduction}


Current physical implementations of quantum processors utilize qubits based on  trapped ions~\cite{Monroe21}, neutral atoms~\cite{LukinNature21}, nuclear spins~\cite{Pla13,Cory97,Chuang97}, topological qubits~\cite{Marcus18}, superconducting circuits~\cite{Arute19}, quantum dots~\cite{Asaad20,Chatterjee21}, semiconductor spin qubits~\cite{Madzik22}, NV centers in diamond~\cite{Pezzagna21} as well as solid-state qubits made from other color centers~\cite{Rugar21}. 
Spin qubits were among the first experimental realizations towards proposed quantum processors due to their long coherence times and available control methods in magnetic resonance experiments~\cite{Cory97,Chuang97}. In order to build quantum devices with spin qubits, a scalable design that provides individual control and detection is needed~\cite{nadj2010spin,pla2012single,yoneda2018quantum,Chatterjee21}.


Universal quantum computing can be achieved with a minimum set of quantum gates that allow for the implementation of arbitrary quantum algorithms~\cite{preskill1998fault}.
A robust implementation of quantum gates combined with error correction codes is the current prescription for fault-tolerant quantum computing~\cite{Gottesman18}.  The creation of high-fidelity single and two-qubit gates remains a challenge in every implementation, especially those involving spin qubits that are spatially localized at the atomic to nanoscales.  At those length scales, the selective control of spin qubits is demanding because of the difficulty in creating strong, localized control fields that affect only the qubits in the volume of interest, while minimizing cross-talk with neighboring regions.

In this work, we show that for an isolated electron system, individual control of spin qubits can be realized using nanomagnets. 
Nanoscale magnets present two key advantages in controlling spin qubits: (1) Unlike collective application of microwaves in magnetic resonance experiments, they allow for the application of highly localized magnetic fields that minimize the effect on neighboring qubits. 
(2) They offer an  extremely energy efficient pathway for the control of qubits. This leverages spintronic methods for energy-efficient manipulation of magnetization through the use of spin-orbit-torque (SOT)~\cite{liu2012spin,pai2012spin,niimi2012giant}, voltage control of magnetic anisotropy (VCMA)~\cite{maruyama2009large,shiota2009voltage,shiota2012induction,grezes2016ultra}, strain mediated voltage control or “straintronic” based methods~\cite{atulasimha2010bennett,cui2015generation,d2016experimental,mathurin2016stress} and other paradigms for voltage control of magnetism~\cite{heron2011electric}. Energy efficiency is achieved through voltage control, rather than current control, thereby avoiding current dissipation losses~\cite{d2018energy} $(I^{2}R)$ associated with the generation of magnetic fields. 
For example, the energy dissipation per bit for VCMA~\cite{wang2015magnetoelectric} and voltage induced strain from a PZT layer is less than 1 fJ and 100 aJ, respectively, making them 100 and 1,000 times more efficient than  state-of-the-art spin-transfer torque (STT) methods~\cite{nowak2016dependence}, which consume $\sim$100 fJ/bit
~\cite{d2018energy}. Thus, the use of VCMA in controlling the magnetization of nanomagnets results in an energy efficient method for controlling qubits. Another interesting candidate is strain-mediated voltage control. 
Prior work has shown that one can use surface acoustic waves to drive a magnetic film at resonance, which emits magnons in a wide frequency band, some of which produce microwaves that drive transitions in NV centers~\cite{labanowski2018voltage}. However, this does not result in coherent rotations of the qubits.  More recently, coherent rotation of single spin qubits in a NV center~\cite{wang2020electrical} by spin-waves propagating adjacent to it has been demonstrated. Nanoscale manipulation of silicon qubits~\cite{pla2012single,Chatterjee21,yoneda2018quantum} including of flying qubits~\cite{nadj2010spin} have been demonstrated. 

Herein we  demonstrate the feasibility of scalable, small footprint, high- fidelity, energy-efficient quantum gates based on VMCA.
Here, we use electron spins with $g$-factor of 2.0 as a model system to simulate qubit dynamics in the presence of a static external field whose magnitude is comparable to the stray field of the nanomagnets.  This intermediate-field regime is considered more challenging due to the more pronounced effects of spatial inhomogeneities (i.e. spatially varying Larmor frequency and axis of quantization) and the lack of rotating-wave approximation.   We also consider control of qubit ensembles located in a finite-size nanoscale volume, where field inhomogeneities degrade gate fidelity when averaged over the volume.


The choice of implementing nanoscale control of spin ensembles in this work is motivated by recent proposals~\cite{meso1,meso2,meso3,meso4,meso5,meso6,meso7,meso8,meso9} for quantum entanglers, {\it bona fide} qubits, quantum sensing and quantum memory. 
In all cases, high fidelity gate operations are needed. 
However, this comes at a cost, as gates implemented by an ensemble of spins distributed over a volume would suffer lower gate fidelity due to field inhomogeneities. This is studied here to ensure that we derive the benefits of spin ensembles while still achieving high gate fidelities.
We shall use the term ``qubit volume'' to refer to the mesoscopic region enclosing the spin ensemble of interest.

\section{Voltage control of nanomagnets to apply control pulses to the Qubit}

\begin{figure*}
\centering
\includegraphics[width=0.8\textwidth]{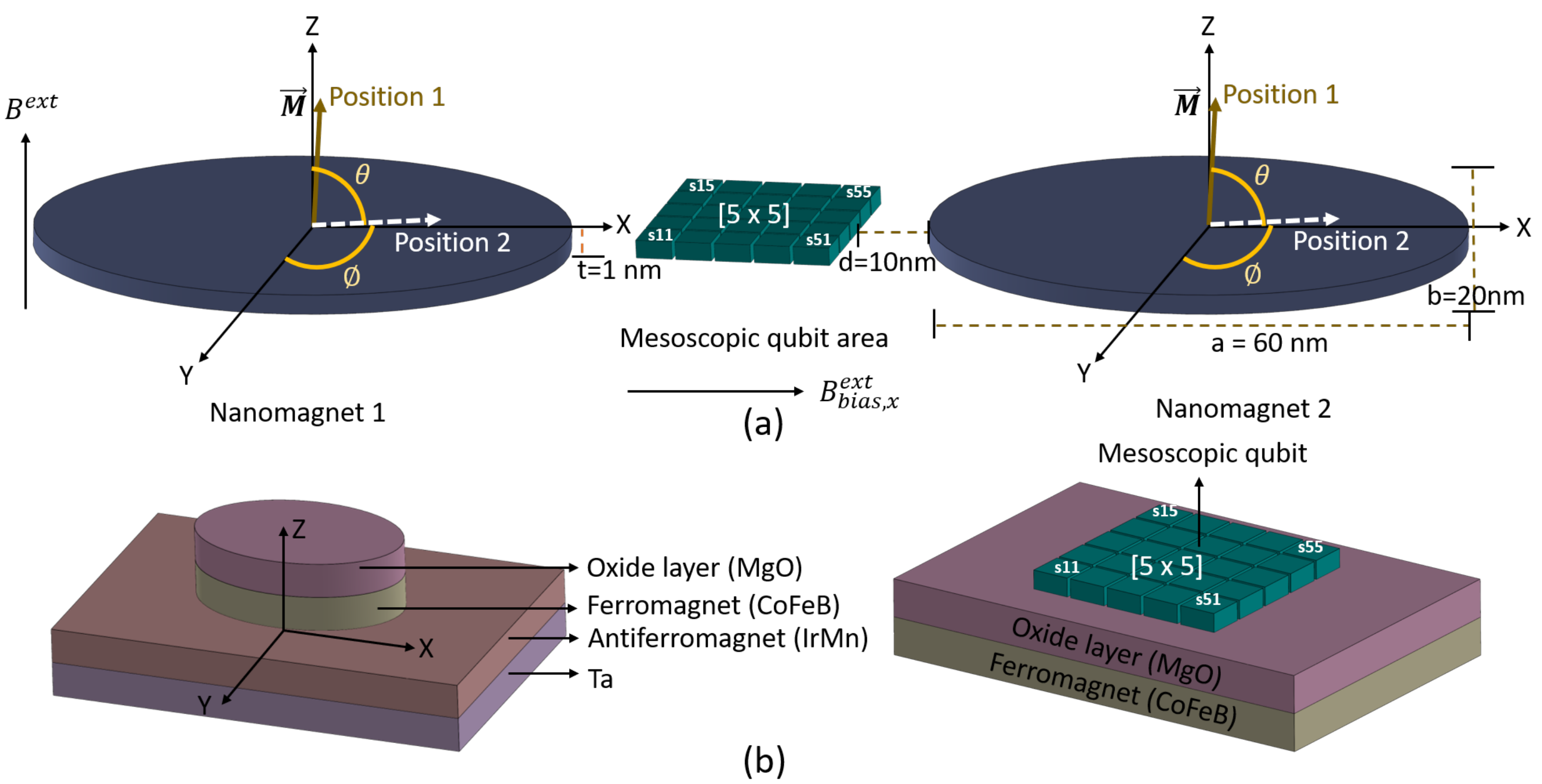}
\caption{(a) Schematic diagram of simulation setup in MuMax3. (b) Schematics of nanomagnet and qubit layers. (Left) Ferromagnet/antiferromagnet interface to create exchange bias field in $+x$ direction in the nanomagnet, so it deterministically oscillates from $+z$ to $+x$ when VCMA is applied in an external bias field of 0.3 T. (Right) FM/Oxide interface to create PMA in the film (magnetized to point along $-z$) below the qubit volume that cancels part of the external magnetic bias field so the effective $+z$ field seen by the qubit produces Larmor precession at 0.5 GHz (or 2 GHz) simulated in this work. (Images are not to scale.)}
\label{fig:simsetup}
\end{figure*}

The magnetization dynamics of the nanomagnets --- as simulated by solving the Landau-Lifshitz-Gilbert (LLG) equation (see Methods section) --- leads to a time varying induced magnetic field in the qubit volume (also assumed to be nanoscale). A schematic diagram of the simulation setup of the qubit volume (5 nm $\times$ 5 nm $\times$ 1 nm) with a nanomagnet in each side is shown in Fig.~\ref{fig:simsetup}(a) and Fig.~\ref{fig:simsetup}(b). The qubit volume consists of a planar array of 25 spins defined as $s_{ij}$; $i$=row number, $j$=column number in each cell so that each spin is separated from its neighbor by 1 nm.

The nanomagnets that drive the Rabi oscillations in these spins by inducing a resonant AC magnetic field due to their magnetization dynamics are elliptical in shape and have length, $a = 60$ nm, width, $b = 20$ nm, and thickness, $t = 1$ nm. Since the qubit volume is near-field as it is very close to the nanomagnets (distance is $\sim$10 nm, which is a fraction of the wavelength $\sim$15 cm), we calculate AC magnetic field at the qubit volume from the magnetostatic field induced by the nanomagnet, which changes as a function of time due to the magnetization dynamics.


The nanomagnets and the qubit volume are assumed to be placed in a uniform external magnetic field pointing along the direction of the  $z$-axis. Due to  perpendicular magnetic anisotropy  (PMA) as well as the global bias magnetic field (along $+z$) the magnetization of the nanomagnets are out-of-plane (and points along $+z$). To alter the magnetizations of the two identical nanomagnets,
PMA is varied through the application of VCMA.

Note that in our case, the VCMA makes the in-plane direction easy, the shape anisotropy due to elliptical nanomagnet shape drives the magnetization to the easy (either $\pm x$) axis of the nanomagnet with equal probability~\cite{Bhattacharya18}. To preferentially orient the magnetization along $+x$, an exchange bias from an underlying antiferromagnet (AFM) can be applied, resulting in a highly localized exchange bias field, $B_{\text{bias}\_x}^{\text{ext}}$ along ($+x$) in each nanomagnet. This exchange bias field can be realized at a ferromagnet/antiferromagnet (e.g. CoFeB/IrMn) interface as shown in Fig.~\ref{fig:simsetup}(b). The rotation of the magnetization to the $+x$-direction due to VCMA induces a magnetic field along $+x$ in the qubit volume, which is located between the two nanomagnets with a distance of 10 nm from each of them indicated as d in Fig.~\ref{fig:simsetup}(a). The magnetization is restored to the $z$-direction when the PMA is increased.

By applying a sinusoidal voltage to the nanomagnets to induce VCMA, a periodic (sinusoidal with higher harmonics due to nonlinear response) magnetic field is induced along $x$-axis which is applied to the spins in the qubit volume and causes Larmor precession of these spins when frequency of this induced field drives the spins at resonant condition for a particular value of the effective magnetic bias field in the $z$-direction (due to the effective global bias magnetic field). 

A ferromagnet/oxide interface below the  qubit volume plays two roles. It creates a PMA in the film, which is magnetized to point along $-z$ axis that cancels part of the external magnetic bias field along $+z$ to produce an effective field which corresponds to the Larmor precession of the spins in the qubit volume at 0.5 GHz (or 2 GHz). The  qubit volume can also be initialized by applying a spin-transfer-torque (STT) current where the MgO acts as a tunnel barrier layer. 

The parameters used in the simulations are listed in Table~\ref{tab:simpar}. The effective bias magnetic field applied in the nanomagnet accounts for an external bias field of 0.3 T along the $z$-direction and the field along the $z$-direction due to PMA.


\begin{table}
\caption{\label{tab:simpar}List of parameters used in the simulation.}
\begin{ruledtabular}
\begin{tabular}{cc}
 \multicolumn{1}{c}{Parameters}&\multicolumn{1}{c}{Value}\\ \hline
Saturation magnetization ($M_s$) &  0.8$\times$10$^6$ A/m \\
Gilbert damping constant ($\alpha$) & 0.1 \\
Exchange stiffness ($A_{\text{ex}}$) & 10$\times$10$^{-12}$ J/m \\
Maximum PMA constant ($K_{u1,\text{max}}$)
& 1.0$\times$10$^6$ J/m$^3$ \\
Exchange bias field ($B_{\text{bias}\_x}^{\text{ext}}$) & 100 mT \\
Nanomagnet thickness ($t$) & 1 nm \\
VCMA frequency ($\nu$) & 500 MHz and 2 GHz \\
VCMA coefficient ($\eta$) & 500 fJ/Vm \\
\end{tabular}
\end{ruledtabular}
\end{table}

\begin{figure}
\centering
\includegraphics[width=\columnwidth]{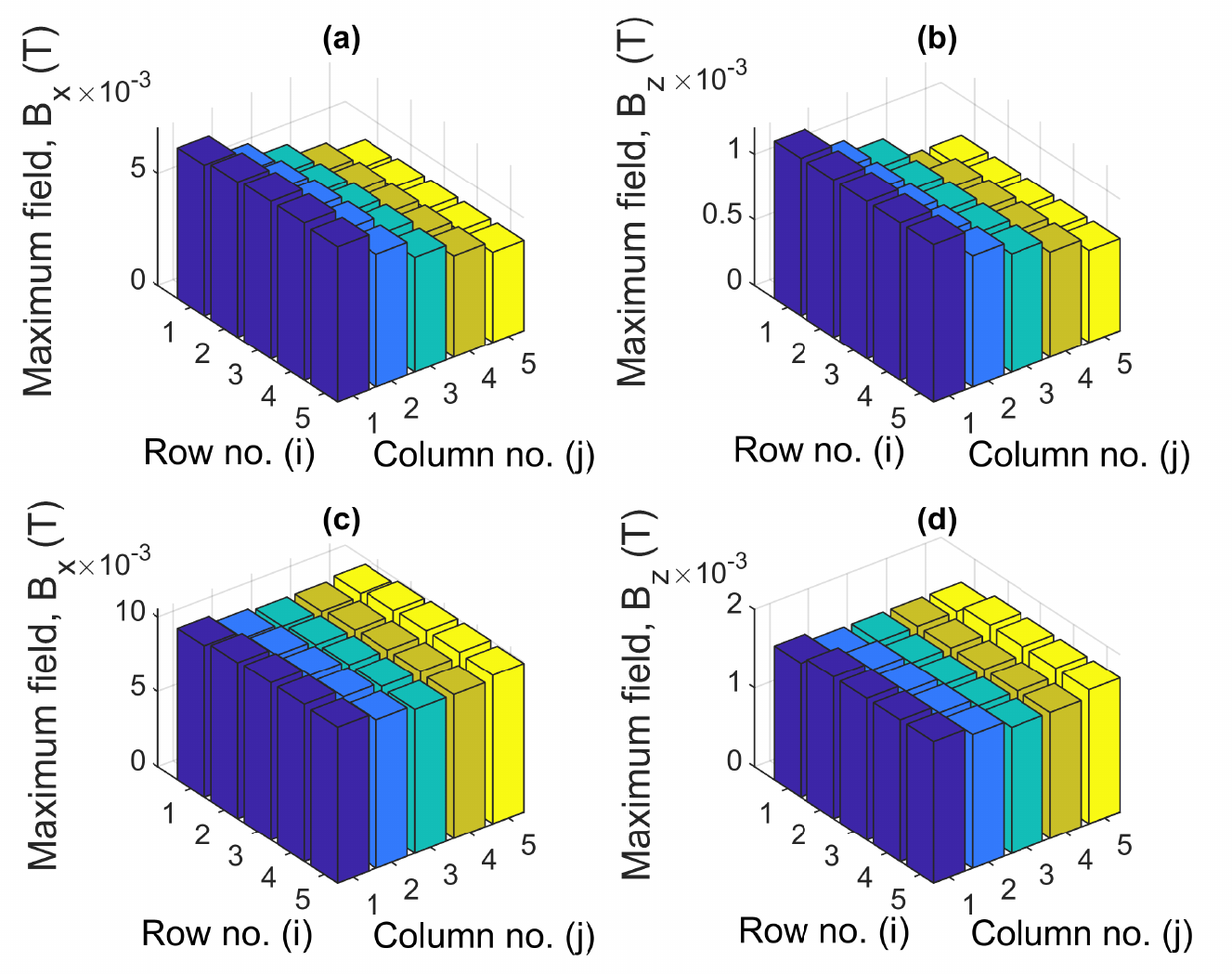}
\caption{Field gradient histogram plots in the  qubit volume (a) Maximum magnetic field of each spin along $x$-axis and (b) along $z$-axis with a single nanomagnet. (c) Maximum magnetic field of each spin along $x$-axis and (d) along $z$-axis with two nanomagnets. }
\label{fig:fieldgrad}
\end{figure}

\section{Induced field profiles}

We simulated and obtained the magnetic field in the qubit volume for two cases: for a single nanomagnet and for two nanomagnets. The histogram plots in Fig.~\ref{fig:fieldgrad} show magnetic field gradients in both the $x$ and $z$ directions with a single nanomagnet and with two nanomagnets. The row and column numbers in the $x$ and $y$ axis correspond to the position of the spin in the qubit volume. In Figs. ~\ref{fig:fieldgrad}(a) and (b), the maximum induced magnetic field along the $x$-axis $B_{\text{max},x}$ and the $z$-axis $B_{\text{max},z}$ are given for each of the 25 cells considered in the qubit volume for the case of a single nanomagnet. The maximum amplitude achieved is 0.007 T and the field gradient is 0.003 T or 42.86 \% in the $x$-direction. This field gradient creates inhomogeneity and leads to low fidelity of quantum gate operations. The simulation result shows a reduced magnetic field gradient and improved amplitude in both $x$ and $z$ directions for two nanomagnets. The maximum amplitude $B_{\text{max},x}$ achieved is 0.011 T which is comparatively higher, and the field gradient is 0.001 T or 9.09 \%, which is comparatively lower than for the case with a single nanomagnet.

\begin{figure*}
\centering
\includegraphics[width=0.8\textwidth]{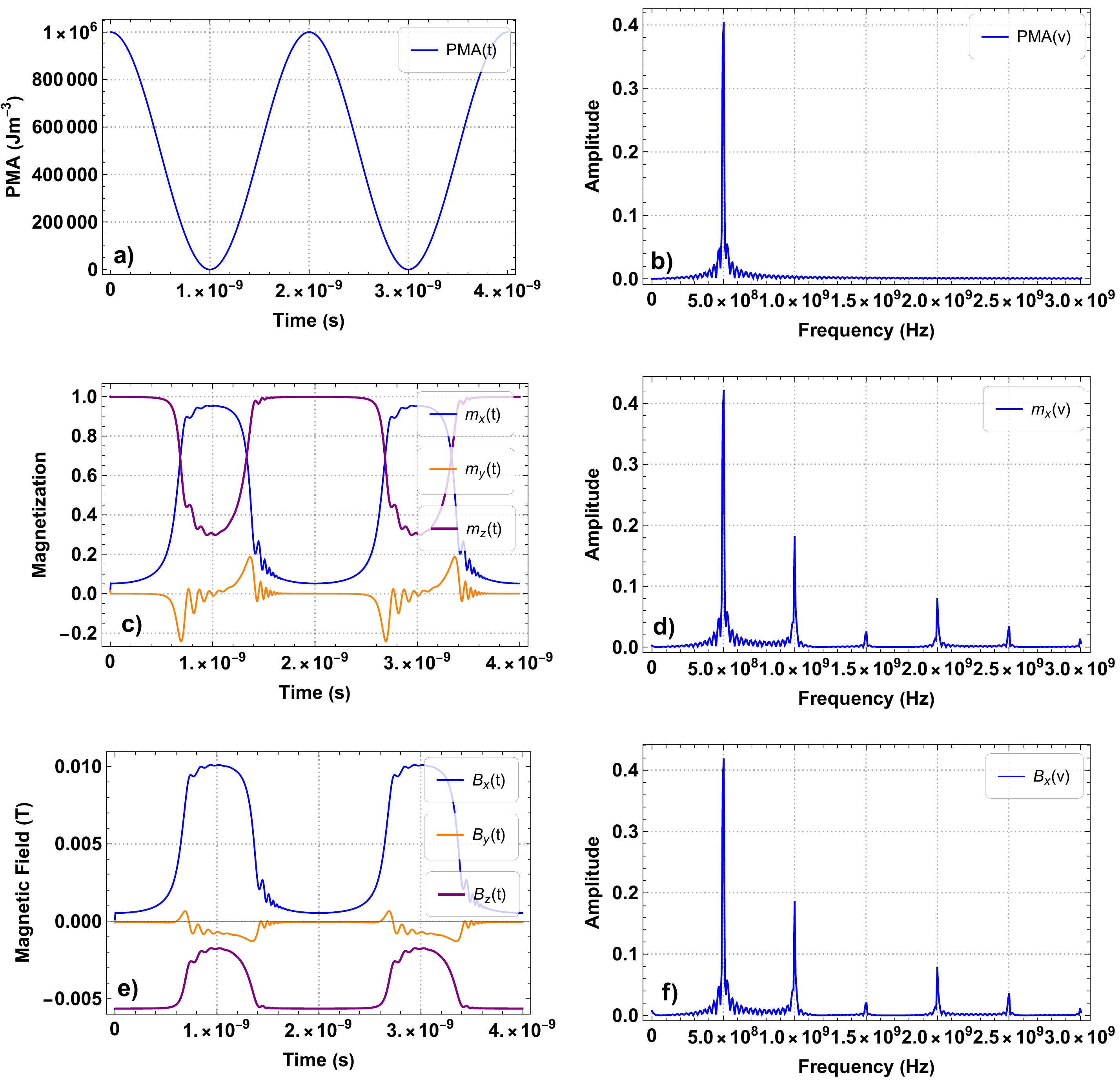}
\caption{Sinusoidal (500 MHz) perpendicular magnetic anisotropy in (a) time and (b) frequency domain. Magnetization ($m_x$, $m_y$, $m_z$) of a nanomagnet in (c) time and (d) frequency domain. Induced magnetic field ($B_x$, $B_y$, $B_z$) in the qubit volume in (e) time and (f) frequency domain.}
\label{fig:500MHzH1}
\end{figure*}

The sinusoidal variation of PMA applied to the nanomagnets through VCMA, VCMA-induced sinusoidal variation of PMA leading to magnetization dynamics and inducing a time varying magnetic field in the qubit volume
are shown in Fig.~\ref{fig:500MHzH1} along with corresponding frequency domain plots. A sinusoidal VCMA of 500 MHz is shown in time domain and frequency domain in Fig.~\ref{fig:500MHzH1}(a) and Fig.~\ref{fig:500MHzH1}(b). The magnetization in the nanomagnet (Fig.~\ref{fig:500MHzH1}(c)) and the induced magnetic field (Fig.~\ref{fig:500MHzH1}(e)) contains higher harmonics (1 GHz, 2 GHz etc.) due to the nonlinear response of the nanomagnet to VCMA as shown in Fig.~\ref{fig:500MHzH1}(d) and Fig.~\ref{fig:500MHzH1}(f). The magnetization of the nanomagnet pointing in the $+z$ axis induces a magnetic field in negative $z$ direction due to the dipole effect.

\begin{figure*}
\centering
\includegraphics[width=\textwidth]{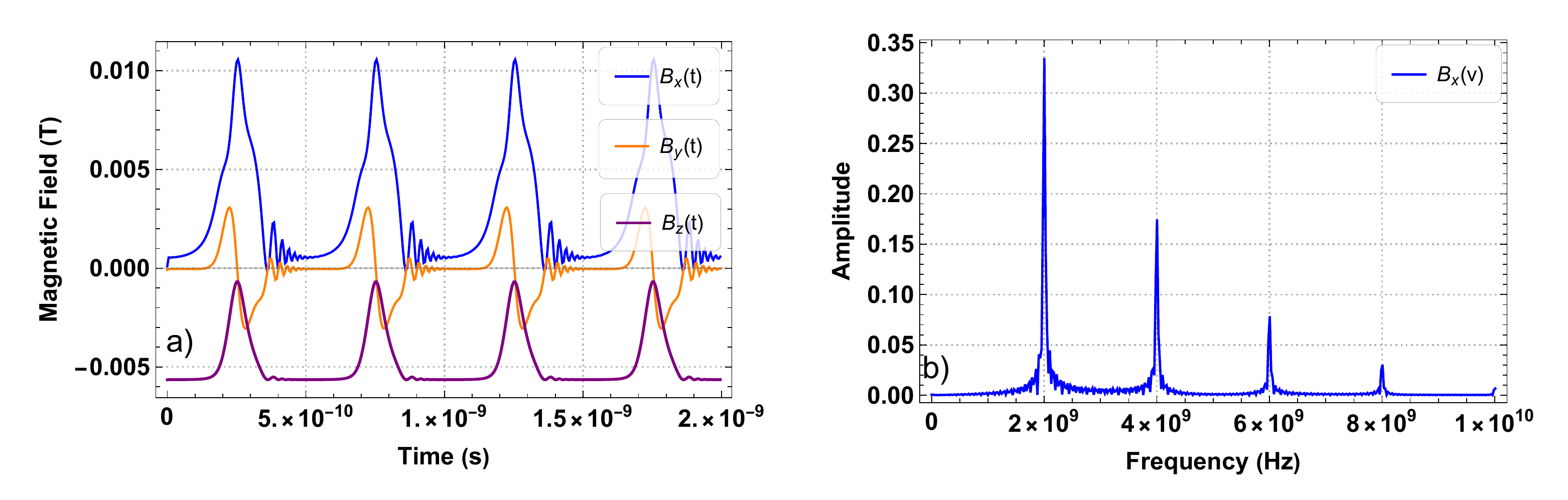}
\caption{Induced magnetic field ($B_x$, $B_y$, $B_z$) in the qubit volume in (a) time and (b) frequency domain.}
\label{fig:2GHzH1}
\end{figure*}

The induced magnetic field ($B_x$, $B_y$, $B_z$) in the qubit volume in response to a 2 GHz sinusoidal VCMA applied in the nanomagnets and its frequency domain plot are shown in Fig.~\ref{fig:2GHzH1}. The $x$-component ($B_x$) contains 2 GHz as well as higher order harmonics such as 4 GHz, 6 GHz etc.

For a single frequency control pulse, perfect gate implementation is possible in theory. It can be shown that as other harmonics add to the control field, reaching the same clean rotations become more challenging and we expect a drop in the gate fidelity as the number of Fourier components increases.  Fourier decomposition is performed for the induced magnetic field at 500~MHz, Fig.~\ref{fig:pulsecomp} panels a, b, and c, and for 2~GHz, in panels d, e, and f. The first 20 Fourier coefficients of each field are plotted in the inset. Since the 2~GHz field has a smaller number of components we expect gate operations with larger fidelities in comparison with the 500 MHz drive. 


\section{Spin dynamics}

We show how to use the induced field of a nanomagnet for  implementation of single-qubit gates on electron-spin qubits.  By describing the evolution of spins, we show that despite the complex nature of the induced field profile, robust implementation of quantum gates is achievable. 

The Hamiltonian, $\mathcal{H}(t)=-\gamma_e\vec{B}(t) \cdot \vec{S}$, for spin interaction with a magnetic field $\vec{B}(t)=\vec{B}_0(t)+\vec{B}_1(t)$ is
\begin{align*}
\mathcal{H}(t)=&-\omega_{1x}(t)\hat{S}_x-\omega_{1y}(t)\hat{S}_y  -(\omega_0+\omega_{1z}(t))\hat{S}_z \\ \nonumber
=& -(\omega_{1x}^{\text{st}}+\omega_{1x}^{\text{var}}(t))\hat{S}_x-(\omega_{1y}^{\text{st}}+\omega_{1y}^{\text{var}}(t))\hat{S}_y\\
&-(\omega_0+\omega_{1z}^{\text{st}}+\omega_{1z}^{\text{var}}(t))\hat{S}_z \\ \nonumber
=& \mathcal{H}_0 + \mathcal{H}_1(t) \nonumber
\end{align*}
where $\omega_0=\gamma_e B_0$ is the angular velocity of electron spins when subjected to the external static field $B_0$, or the Larmor frequency, $\gamma_e$ is the gyromagnetic ratio of the electron and $\omega_{1\alpha}(t)= \gamma_e \vec{B}_1(t) \cdot \hat{\alpha}$ for $\hat{\alpha} \in \{ \hat{x},\hat{y},\hat{z} \}$, is proportional to the strength of the control field in each direction. The time-independent portion of the Hamiltonian $\mathcal{H}_0=-\omega_{1x}^{\text{st}}\hat{S}_x-\omega_{1y}^{\text{st}}\hat{S}_y -(\omega_0+\omega_{1z}^{\text{st}})\hat{S}_z $ includes the external field $B_0$ and the time-independent part of the nanomagnet induced field $B_{1}^{\text{st}}$. This is a result of the bias field applied to fix the rotation direction of the magnetization vector in the nanomagnet. $\mathcal{H}_1(t)=-\omega_{1x}^{\text{var}}(t)\hat{S}_x-\omega_{1y}^{\text{var}}(t)\hat{S}_y -\omega_{1z}^{\text{var}}(t)\hat{S}_z$ represents the time-dependent part of the induced magnetic field, which is used to control qubits, in place of RF or microwave pulses. 

\begin{figure*}[tbh]
\centering
\includegraphics[width=\textwidth]{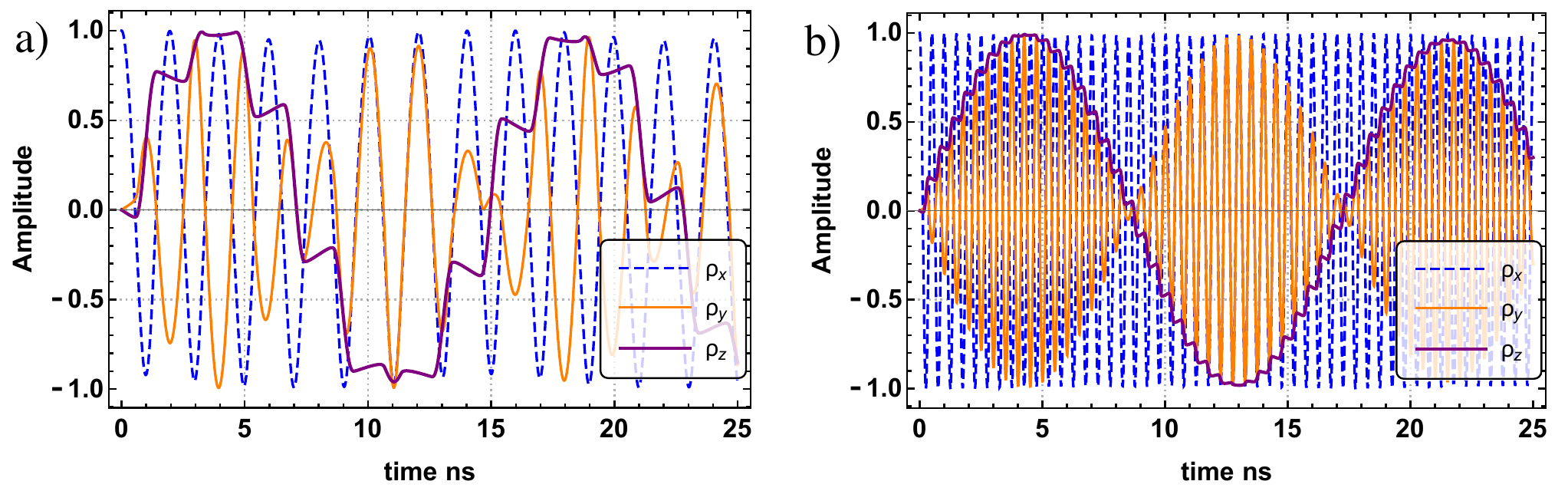}
\caption{Evolution of spins in the lab frame for nanomagnet induced field with 500 MHz drive (Left) and 2 GHz drive (Right). Average evolution of 25 spins initialized along the $x$, $y$, and $z$ directions are projected on $x$, $z$, $y$ axes, respectively. The effective $\pi/2)_{x}$ rotation happens at 3.871$\pm0.001$ ns and 4.498$\pm0.001$ ns in 500 MHz and 2 GHz drives, respectively. Spins rotate a $\pi$ Radian along the x-axis at 7.882$\pm0.001$ ns and 8.998$\pm0.001$ ns in 500 MHz and 2 GHz frequencies, respectively. }
\label{fig:contpulses}
\end{figure*}

Spin dynamics in the lab frame, is described with the Liouville von-Neumann equation, Eq.~(\ref{eq:lvnR}), 
with a unitary propagator defined as 
\begin{equation}\label{eq:U1}
U_1=\tau \exp\left\{-i \int_0^t \left[ \mathcal{H}_0+\mathcal{H}_1(t') \right] dt'\right\} 
\end{equation}
where $\tau$ is the Dyson time ordering operator. 

The induced magnetic field of nanomagnet has a pronounced static field along $x$ and $z$ directions. These time-independent field components are part of the $\H_0$ Hamiltonian and as a result, the spins precess around an effective field defined by these fields, which is in the $x-z$ plane, slightly deviating from the $z$-axis. The angular velocity for this precession is $$ \omega_r=\sqrt{(\omega_{1x}^{\text{st}})^2+(\omega_{1z}^{\text{st}}+\omega_0)^2}.$$
Time-independent components of the induced field, $\omega_{1x}^{\text{st}}$ and $\omega_{1z}^{\text{st}}$ are evaluated using the time average of field components.
Considering that $\omega_r$ should be in resonance with the drive frequency of the nanomagnet,   the amplitude of the static external field is chosen such that $\omega_0$ satisfies this equation. 

The unitary propagator is evaluated for the continuous application of drive voltage using Eq.~(\ref{eq:U1}). Spin evolution shows the step-wise rotation of spins modulated with the rotation along the effective field with angular velocity $\omega_r$. Depiction of spin dynamics by initializing one electron spin along the $x$, $y$, and $z$ axes and projecting it on the $x$, $z$, and $y$ axes after its rotation. Fig.~\ref{fig:contpulses} shows the average observed signal for both drive frequencies at 500 MHz and 2 GHz. Since the largest  time-dependent field component is along $x$, as we apply these pulse segments in resonance with the Larmor frequency of electron spin, 
we observe $x$ rotations. As expected, the density matrix initialized along the $x$-axis only precess around the effective field with no change in time.  Density matrices initialized in the $y-z$ plane, on the other hand, are affected by the $x$ rotations.   

These results are similar to the spin rotations in the traditional magnetic resonance experiments where spin control is implemented using RF pulses in resonance with the Larmor frequency of the spins in the external magnetic field. In the 2 GHz drive example, since the spin rotation happens in smaller steps, there is a smoother transition and we have more control for single-qubit gate implementation. The $X/2$ gate which is a $\pi/2$ rotation along the $x$-axis can be achieved by stopping the drive when $\rho_y$ rotates to -$z$, or equivalently when $\rho_z$ rotates to $y$. This rotation happens at 3.872$\pm 0.001$ ns for 500 MHz case and at 4.498$\pm 0.001$ ns in the case of 2 GHz drive. A sudden change of drive voltage, especially mid pulse, will cause oscillatory residual magnetic fields a.k.a.  ringing effect. Ideally, we would like to implement gates that last an integer number of pulse segments to minimize the ringing effect. Rotations along the $y$-axis, or any other orientation in the $x-y$ plane, are implemented by shifting the phase of these $X$ pulses, which is done by applying delays before the start of the pulse train. The $X$ gate is achieved at 7.882$\pm 0.001$ ns for 500 MHz and at 8.998$\pm 0.001$ ns for 2 GHz. Table~\ref{tab:gates} shows the field profile of two rotations necessary for the implementation of Clifford gates, for the 2 GHz drive case.

\begin{table*} [ht]
\centering
  \caption{Implementation of quantum gates using nanomagnets at 2GHz drive} \label{tab:gates}
\begin{tblr}{
  colspec={|Q[m,c,40mm]|Q[m,c,50mm]|Q[m,c,60mm]|},
  hline{1,2,7,12,17},
  cell{2}{2} = {r=5}{m}, 
  cell{2}{3} = {r=5}{m},
  cell{7}{2} = {r=5}{m},
  cell{7}{3} = {r=5}{m},
  cell{12}{2} = {r=5}{m},
  cell{12}{3} = {r=5}{m},
}
 Gate & Induced Field Profile & Fidelity Map Volume Average 25 spins \\ 
\textbf{X/2} & \hspace{-0.25\textwidth}{\parbox[c]{1em}{
\includegraphics[width=0.5\columnwidth]{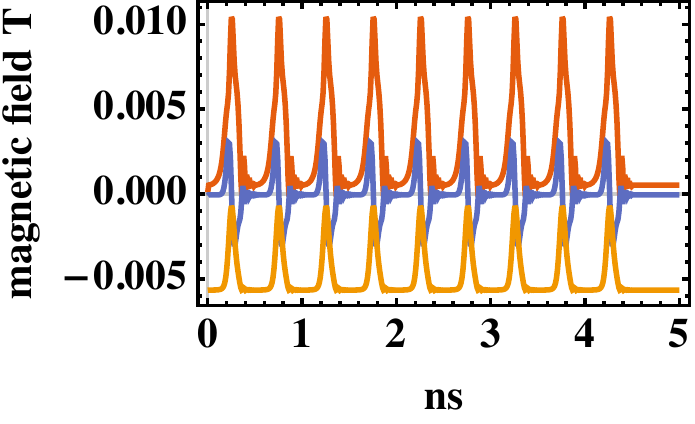}}} & \hspace{-0.25\textwidth}{\parbox[c]{1em}{
\includegraphics[width=2in]{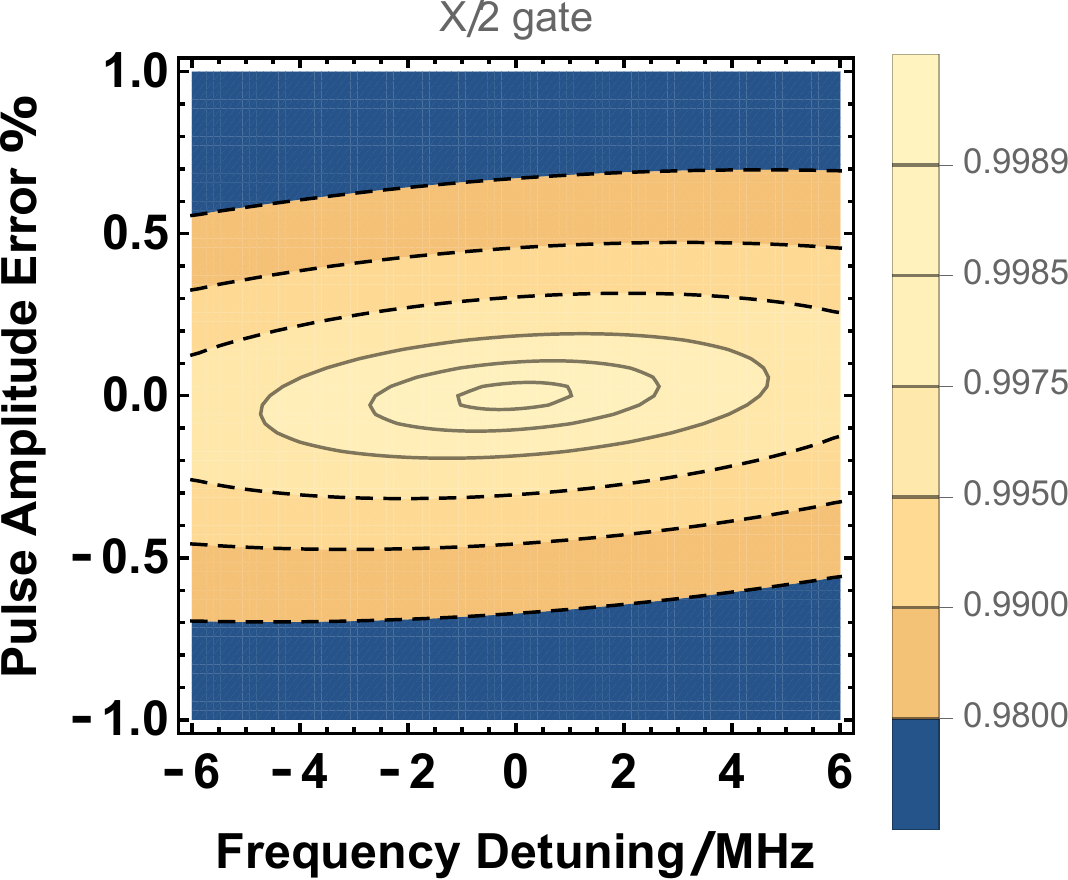}}}  \\
  &  &  & \\
 Ideal Hamiltonian $\mathcal{H}=\exp\left(\frac{i}{2}\frac{\pi}{2} \sigma_{\text{x}}\right)$  &  &  \\
Central spin fidelity:  0.9997 $\pm 10^{-4}$  &  &  \\ 
Fidelity averaged over lattice sites: 0.9990$\pm 10^{-4}$ & & \\
 \textbf{X}  & \hspace{-0.25\textwidth}{\parbox[c]{1em}{
\includegraphics[width=0.5\columnwidth]{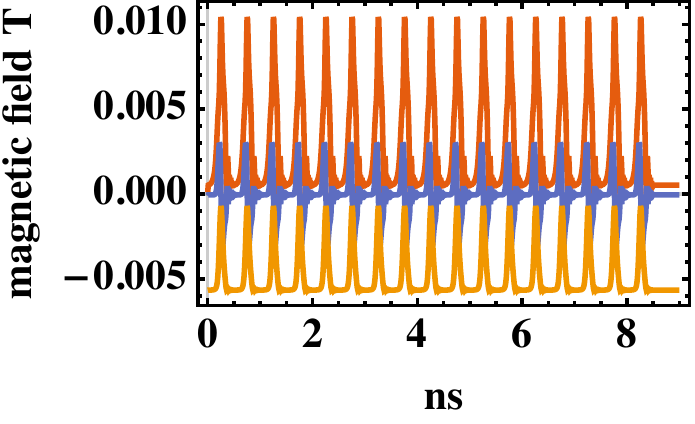}}} & \hspace{-0.25\textwidth}{\parbox[c]{1em}{
\includegraphics[width=2in]{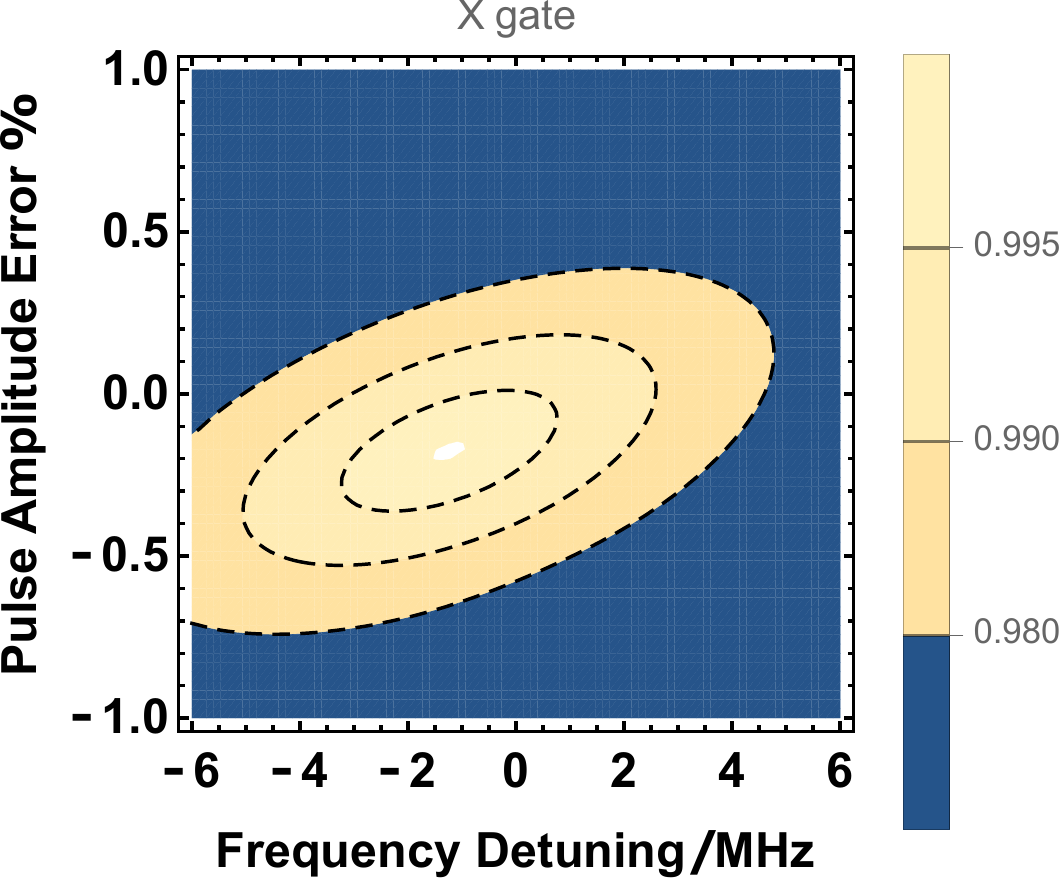}}} \\
  & & & \\
 Ideal Hamiltonian $\mathcal{H}=\exp\left(\frac{i}{2}\pi \sigma_{\text{x}}\right)$ &  &  \\
Central spin fidelity:  0.9987 $\pm 10^{-4}$ &  &  \\ 
Fidelity averaged over lattice sites: 0.9952$\pm 10^{-4}$ & &\\
 \textbf{X Knill CP}  & \hspace{-0.25\textwidth}{\parbox[c]{1em}{
\includegraphics[width=0.5\columnwidth]{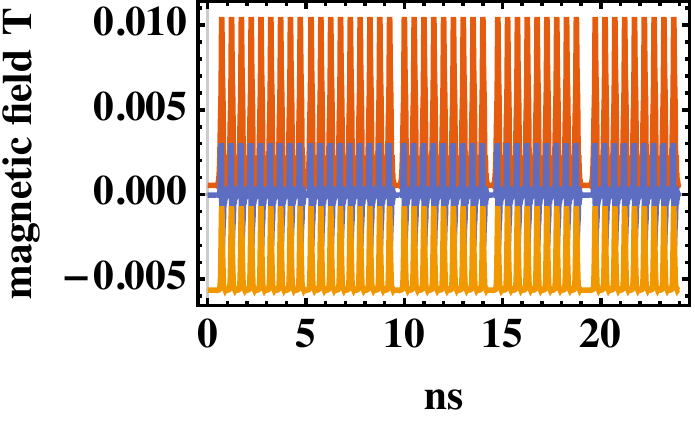}}} & \hspace{-0.25\textwidth}{\parbox[c]{1em}{
\includegraphics[width=2in]{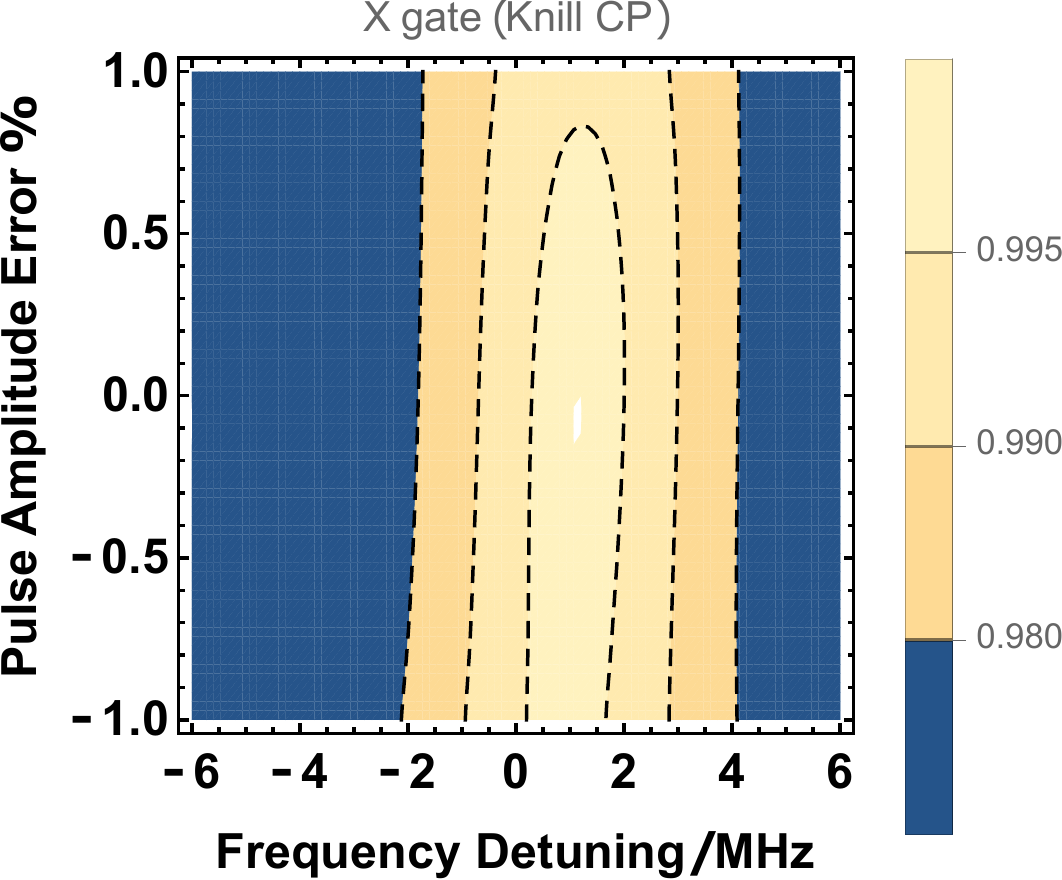}}}  \\
  & & & \\
 Ideal Hamiltonian $\mathcal{H}=\exp\left(\frac{i}{2}\pi \sigma_{\text{x}}\right)$ &  & \\
Central spin fidelity:  0.9993 $\pm 10^{-4}$ &  &  \\
Fidelity averaged over lattice sites: 0.9979$\pm 10^{-4}$ & &
\end{tblr}
\end{table*}

In quantum information processing, error correction codes are used to prevent the loss of quantum information due to imperfections of quantum control. The idea is that if the gates are implemented with enough fidelity or with acceptable error probability per gate (or simply error per gate, EPG), they can be effectively used for quantum information processing. This threshold for EPG is determined by further assumptions of the error model and device parameters and is often between $10^{-6}$ and $3\times 10^{-3}$ (see Refs.~\cite{Knill05,Veldhorst14,Morton16,Chatterjee21}). The typical value used as a threshold for experimental implementation of quantum computers is the EPG of less than $10^{-4}$ (see Ref.~\cite{gottesman1997stabilizer}). 
In the case of two-level systems, average gate fidelity can be computed by comparing the ideal and noisy implementation of unitary maps~\cite{BOWDREY02}. For a general, linear, and trace-preserving map $\mathcal{M}$, and its ideal counterpart unitary $U$, the gate fidelity averaged over initial states is defined in terms of the Hilbert-Schmidt inner product   
\begin{equation}
\label{eq:fidelity}
\bar{F}=\frac{1}{2}+\frac{1}{3}\sum_{j=x,y,z}\mbox{Tr}\left[U\frac{\sigma_j}{2}U^{\dagger}\mathcal{M}\left(\frac{\sigma_j}{2}\right) \right].
\end{equation}
As indicated in Table~\ref{tab:gates}, the fidelity of the $X/2$ gate implemented by  nanomagnet, is well above the required EPG threshold at $99.97\pm 0.01\%$.  This fidelity value applies to a single spin located at the center of the qubit volume.  When the qubit has a finite volume, the fidelity will degrade due to field inhomogeneities over the volume. 
To assess the impact of  inhomogeneities, we view the fidelity as a function of position (i.e. $\bar{F} \equiv \bar{F}(\mathbf{r})$) and average over the lattice sites:
$$ \langle \bar{F} \rangle_{\Omega} = \frac{1}{N} \sum_{\mathbf{r}_i \in \Omega}^N \bar{F}(\mathbf{r}_i) $$
where $\Omega$ is the set of lattice points, $\bar{F}(\mathbf{r}_i)$ is the fidelity of the gate at lattice site $\mathbf{r}_i \in \Omega$ and $N=|\Omega|$ is the number of lattice sites. For the $X/2$ gate the volume averaged fidelity drops to $99.90\pm 0.01\%$.
The fidelity for the longer $X$ gate ($99.87\pm 0.01\%$ at the central spin), on the other hand, falls slightly short of the required threshold. Longer pulses may be improved by using composite pulses that compensate locally for errors in the rotation angles. This is a well-established method in nuclear magnetic resonance for the design of robust pulses~\cite{LEVITT86,Ryan10}. The so-called Knill pulse is a composite $\pi$ pulse designed specifically to be robust against frequency offset and pulse amplitude fluctuation errors. It is the symmetric combination of the following five pulses
$$ \pi_0^{\text{ideal}}(\frac{\pi}{3})_z=(\pi)_{\frac{\pi}{6}}(\pi)_0(\pi)_{\frac{\pi}{2}}(\pi)_0(\pi)_{\frac{\pi}{6}}.$$ 
An $X$ gate implemented with Knill composite pulse shows significant improvement and reaches an average gate fidelity of 0.9993$\pm 10^{-4}$.
The lattice-averaged fidelity of the composite pulse,
0.9979 $\pm 10^{-4}$, is only slightly below threshold.

Lattice averaging was done on a 2D planar lattice $\Omega$ consisting of $N=25$ spins separated 1 nm apart. As expected, performance degrades compared to the case of 1 spin.
This merely reflects the well-known fact that it is impossible to be perfectly on  resonance with all the spins simultaneously. The  $B_1$ inhomogeneity thus degrades the average gate fidelity. Although the resulting fidelities dip below the threshold for acceptable error rates, they remain within $0.1\%$ of it. A map of average gate fidelity for spin-lattice, with frequency detuning of $\pm 0.3\%$ and pulse amplitude errors of $\pm 1\%$ is plotted in table~\ref{tab:gates}. The results show high fidelity regions that depict the robustness of our gates. 

\section{Conclusion}

In conclusion, the use of nanoscale magnets allows the production of highly localized AC magnetic fields to implement single qubit gates.  Despite the highly nonlinear response of the magnetization of the nanoscale magnets to an electric field, we are able to achieve high single qubit gate fidelity through the appropriate use of robust composite pulses. 
The significance of using nanomagnets for quantum control is that we can achieve local control over spin qubits. The power required to oscillate nanomagnets at 500 MHz and 2 GHz by voltage control is lower than that required to power a coil by current control.
We note that no attempt was made here to optimize the results.  With some effort, the geometric arrangement could be improved, for example, to increase the homogeneity and/or minimize the stray field affecting neighboring qubits.  The use of composite pulses allows for qubit control that is robust with respect to field inhomogeneity (Table~\ref{tab:gates}). Combining the burgeoning spintronic field of energy efficient voltage control of magnetism with quantum computing with robust spin qubits, will stimulate
further experiments in energy efficient, robust quantum computing devices at temperatures of a few K.  

\section{Methods}
\subsection{Micromagnetics}

The simulations of the magnetization dynamics in the nanomagnets are performed by solving the LLG equation using a micromagnetic framework (MuMax3~\cite{Vansteenkiste14}) 
\begin{equation}
\frac{d \vec{m}}{d t} = - \frac{ \gamma \vec{m} \times \vec{H}_{\text{eff}} }{(1+\alpha^2)}  - \frac{\alpha \gamma \vec{m} \times (\vec{m} \times \vec{H}_{\text{eff}})} {(1+\alpha^2)} 
\end{equation}
Here, $\alpha$ is the Gilbert damping coefficient, $\gamma$ is the gyromagnetic ratio, $\vec{m}=\frac{\vec{M}}{M_s}$ is the normalized magnetization, where $\vec{M}$ is the magnetization and $M_s$ is the saturation magnetization.
The effective magnetic field, $\vec{H}_{\text{eff}}$ in this case consists of the fields due to the exchange interaction, uniaxial anisotropy of the nanomagnets, and the demagnetizing field.
$$ \vec{H}_{\text{eff}} = \vec{H}_{\text{an}} + \vec{H}_{\text{ex}} + \vec{H}_{\text{d}} $$
where $\vec{H}_{\text{an}}$ is the effective field due to the uniaxial perpendicular magnetic anisotropy (PMA) which can be modulated using voltage control of magnetic anisotropy (VCMA), $\vec{H}_{\text{ex}}$ is the effective field due to Heisenberg exchange coupling and $\vec{H}_{\text{d}}$ is the field due to the demagnetization energy (shape anisotropy). 

The effective field due to the perpendicular magnetic anisotropy, $\vec{H}_{\text{an}}$ is given as:
$$ \vec{H}_{\text{an}} = \frac{ 2 K_{u1} }{ \mu_0 M_s} (\vec{z} \cdot \vec{m}) \vec{z}. $$
Here, the first order uniaxial anisotropy constant is $K_{u1}$, the magnetic permeability of free space is $\mu_0$, and $\vec{z}$ is the unit vector corresponding to the anisotropy direction. 

While PMA is created from the interaction between the ferromagnet's hybridized $d_{xz}$ and oxygen's $p_z$ orbital at a ferromagnet/oxide interface~\cite{Yang11}, by the application of voltage pulse, the interface electron density as well as perpendicular anisotropy can be changed~\cite{Niranjan10}. This phenomenon is called VCMA~\cite{Amiri12,Wang12,li_lee_razavi_18}. 

The cell sizes are chosen to be 1 nm$^3$, so that all dimensions are well within the limit of ferromagnetic exchange length calculated by $\sqrt{2A_{\text{ex}}/\mu_0 M_s^2} \approx 4.99$ nm.

\subsection{Quantum control with periodic, polychromatic, inhomogeneous field}

Spin dynamics in magnetic resonance experiments is governed by the time-dependant part of the Hamiltonian resulting from the application of r.f. pulses. Consider the Zeeman interaction between spin $\vec{S}$ and external static field ($\vec{B}_0=B_0 \hat{z}=(\omega_0/\gamma)\hat{z}$, where $\gamma$: gyromagnetic ratio, $g\mu_B/\hbar$ for electrons or $g_n\mu_N/\hbar$ for nuclei) and also time-dependent r.f. fields $$\vec{B}_1(t) = (\omega_x(t)/\gamma) \hat{x} + (\omega_y(t)/\gamma) \hat{y}+ (\omega_z(t)/\gamma) \hat{z}$$  
is:
\begin{equation*}
\begin{split}
\mathcal{H}(t) &= -\gamma \left(\vec{B}_0 + \vec{B}_1(t) \right) \cdot \vec{S}  \\
&= -\omega_x(t) \hat{S}_x - \omega_y(t) \hat{S}_y  - (\omega_0 + \omega_z(t) ) \hat{S}_z \\
&= \mathcal{H}_0 + \mathcal{H}_1(t)   
\end{split}
\end{equation*}
where
$$ \mathcal{H}_0 = -\omega_0 \hat{S}_z, \quad  \mathcal{H}_1(t) = -(\omega_x(t) \hat{S}_x + \omega_y(t) \hat{S}_y + \omega_z(t)  \hat{S}_z).  $$
Denoting operators transformed to the rotating frame by a tilde, we write:
$$ \tilde{\mathcal{H}} (t) = e^{ i \omega_r \hat{S}_z t } \mathcal{H}(t) e^{- i \omega_r \hat{S}_z t} $$
$$ \tilde{\rho}(t) = e^{ i \omega_r \hat{S}_z t } \rho(t) e^{- i \omega_r \hat{S}_z t} $$
By differentiating the latter expression  with respect to time, we find the evolution of density matrix in the interaction representation, a.k.a. Liouville von-Neumann equation:
\begin{equation}\label{eq:lvnR}
\frac{\partial\tilde{\rho}}{\partial t} = - i [ \tilde{\mathcal{H}}(t), \tilde{\rho}(t) ] \qquad 
\end{equation}
where
\begin{align} 
\tilde{\mathcal{H}}(t) =& \tilde{\mathcal{H}}_0 + \tilde{\mathcal{H}}_1(t) - \omega_r \hat{S}_z  \nonumber
\\
=& e^{ i \omega_r \hat{S}_z t}  \left[- \omega_x (t) \hat{S}_x -  \omega_y(t) \hat{S}_y \right] e^{ - i \omega_r \hat{S}_z t}+\\ 
& \qquad (-\omega_0 - \omega_z(t) + \omega_r) \hat{S}_z  \label{eq:HR}
\end{align}
The solution to Eq.~(\ref{eq:lvnR}) is given in terms of time-ordered exponentials:
$$ \tilde{\rho}(t) = \mathcal{T} e^{ - i \int_0 ^t \tilde{\mathcal{H}}(t) dt } \tilde{\rho}(0) \mathcal{T} e^{ i \int_0 ^t \tilde{\mathcal{H}}(t) dt } .$$
Consider a single frequency r.f. pulse $$\omega_x(t) = w_1 \cos(\omega_r t + \phi(t) ), \quad \omega_y(t) = w_1 \sin(\omega_r t + \phi(t) )$$ and let $\omega_z(t)=0$ for simplicity.
At the resonance condition  $\omega_0=\omega_r$, the spin evolution in the interaction representation is described with 
\begin{multline} \tilde{\mathcal{H}}(t)  = e^{ i \omega_0 \hat{S}_z t}  w_1 [ \cos(\omega_0 t + \phi(t) ) \hat{S}_x \\- \sin(\omega_0 t + \phi(t) ) \hat{S}_y  ] e^{ - i \omega_0 \hat{S}_z t}.
\end{multline}
This is a  rotation with the Rabi frequency  $w_1$. We can extend the analogy for a periodic control field with period $T$, containing multiple frequencies, expressed as a sum over Fourier components
$$ \omega_x(t) = \sum_{n_x=-N}^N  c_{x}[n_x] e^{ i 2\pi n_x t/T}. $$
Here $f_0=1/T$ is the {\it fundamental frequency} and the coefficient in the Fourier space are defined as 
$$c[n] = \frac{1}{T} \int_{0}^T \omega(t) e^{ - i 2 \pi n t/T} dt.$$
Similar expressions also exist for $y$ and $z$.
Substitution into Eq.~(\ref{eq:HR}) gives:
\begin{align*} \tilde{\mathcal{H}}(t)
=& e^{ i \omega_r \hat{S}_z t} \left[ \omega_x (t) \hat{S}_x + \omega_y (t) \hat{S}_y \right] e^{ - i \omega_r \hat{S}_z t}  \\
& \quad + (\omega_0  +\omega_z(t) - \omega_r) \hat{S}_z \\
=& \sum_{n=-N}^N  e^{ i \omega_r \hat{S}_z t} \left( c_x[n] \hat{S}_x + c_y[n] \hat{S}_y \right) e^{ i 2\pi n t/T} e^{ -i \omega_r \hat{S}_z t } \\
+& (\omega_0 + c_z[n] e^{ i 2\pi n t/T} - \omega_r) \hat{S}_z
\end{align*}
By setting the fundamental frequency $2\pi f_0 = \omega_r = \omega_0$ the control Hamiltonian becomes:
\begin{align*} \tilde{\mathcal{H}}(t)
= \sum_{n=-N}^N & e^{ i \omega_0 \hat{S}_z t} \left( c_x[n] \hat{S}_x + c_y[n] \hat{S}_y \right) e^{ i \omega_0 n t} e^{ -i \omega_0 \hat{S}_z t } \\+& c_z[n] e^{ i \omega_0 n t} \hat{S}_z
\end{align*}
Using the ladder operators, $\hat{S}_+ = \hat{S}_x + i \hat{S}_y$ and $\hat{S}_- = \hat{S}_x - i \hat{S}_y$, and $[\hat{S}_z, \hat{S}_\pm] = \pm \hat{S}_\pm$ we have
$$ e^{ i \omega_0 \hat{S}_z t} \hat{S}_{\pm} e^{ - i \omega_0 \hat{S}_z t} = \hat{S}_{\pm} + [\hat{S}_z, \hat{S}_{\pm}] i \omega_0 t + \dots = \hat{S}_{\pm} e^{  \pm i \omega_0  t}, $$
and the Hamiltonian becomes:
\begin{align}
\tilde{\mathcal{H}}(t)
= \sum_{n=-N}^N &  \hat{S}_+ e^{ i \omega_0 (n+ 1) t} c_+[n] + \hat{S}_- e^{ i \omega_0 (n-1) t} c_-[n] \nonumber\\
+& c_z[n] e^{ i \omega_0 n t} \hat{S}_z \label{eq:HFourier}
\end{align}
where we used the shorthand notation:
$$ c_+[n] := \frac{c_x[n]}{2} + \frac{c_y[n]}{2i}, \qquad c_-[n] := \frac{c_x[n]}{2} - \frac{c_y[n]}{2i}. $$
The real parameters $c_{\alpha}[n]$ are the components of the control field oscillating at the Larmor frequency $\omega_0$. The component $c_z[0]$ is the time-average of the field ($z$ component). If the $z$-component is sinusoidal, it has no d.c. component and $c_z[0]=0$.    If there is a d.c. offset (nonzero background field), this will cause a shift in the resonance frequency away from $\omega_0$ by the amount $c_z[0]$.
\begin{figure*}[htb]
\centering
\includegraphics[width=0.95\textwidth]{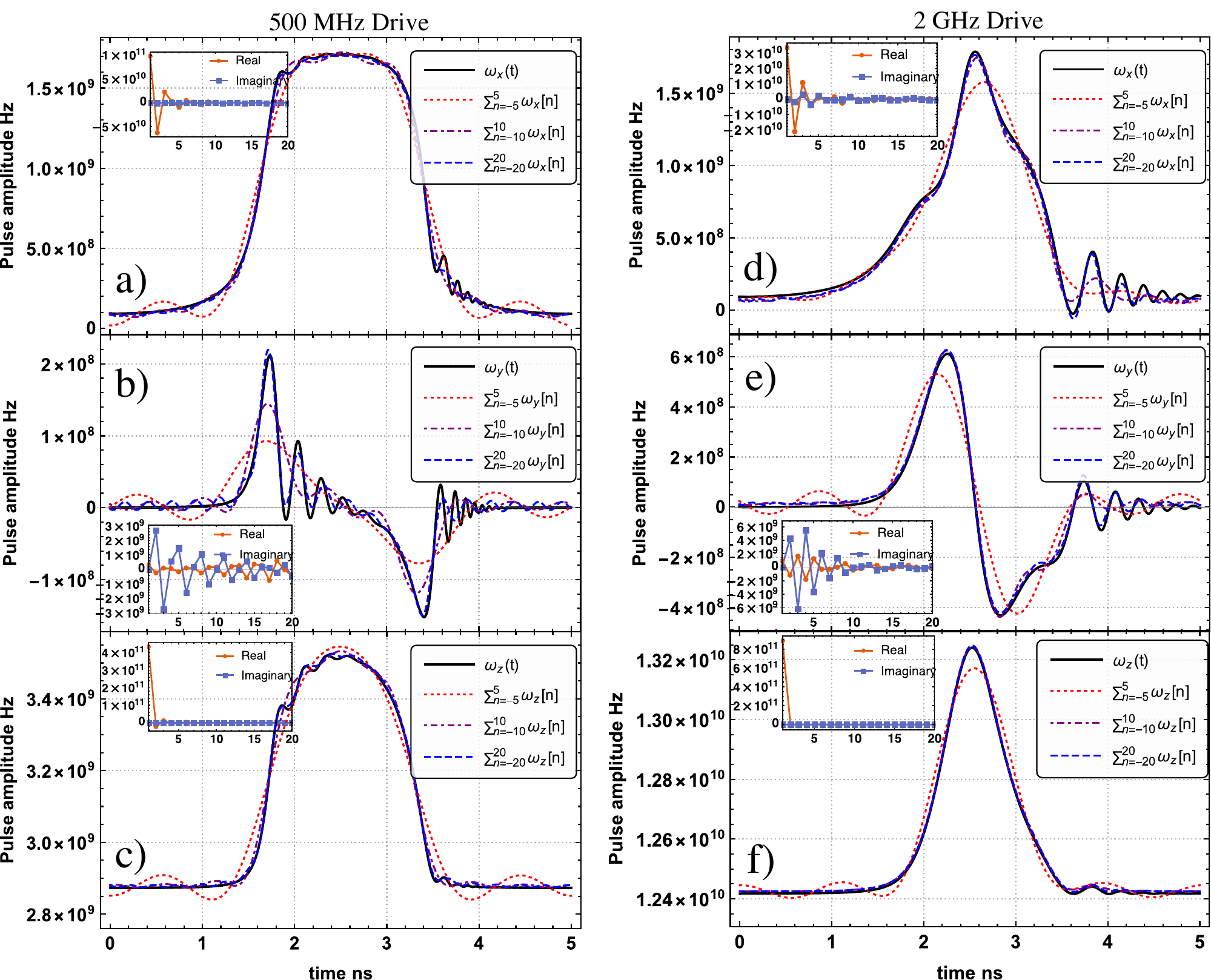}
\caption{$x$, $y$, and $z$ component of the induced magnetic field with 500 MHz drive, are indicated in panels \textbf{a}, \textbf{b}, and \textbf{c} respectively. Panels \textbf{d}, \textbf{e}, and \textbf{f} are showing the results for 2 GHz drive. In each case Fourier components with $N=5,10,20$ are used to reconstruct the original field profile. }
\label{fig:pulsecomp}
\end{figure*}

The $x$, $y$ and $z$ components of one control pulse segment, induced by the nanomagnet at the 500~MHz and 2~GHz drive frequencies are shown in Fig.~\ref{fig:pulsecomp}. For each direction $\alpha$, Fourier components $\omega_{\alpha}[n]=c_{\alpha} \e^{i2\pi n_{\alpha} t/T}$, are added to reconstruct the original time-domain field profile. Notice that by adding  higher number of Fourier components, a better approximation of the induced field is achieved. These components may be used in Eq.~(\ref{eq:HFourier}) to evaluate the unitary propagator for the periodic control field.  From this we conclude that although the presence of harmonics (Fig.~\ref{fig:pulsecomp}) in the control field introduces significant deviations from a sinusoid shape, the presence of an external field comparable to the peak $B_1$ field is sufficient to average away these components and yield a high gate fidelity.  Normally, the rotating wave approximation is only applied in the limit of high fields.

\bibliographystyle{naturemag}
\bibliography{nanomag.bib}

\begin{thebibliography}{10}
\expandafter\ifx\csname url\endcsname\relax
  \def\url#1{\texttt{#1}}\fi
\expandafter\ifx\csname urlprefix\endcsname\relax\def\urlprefix{URL }\fi
\providecommand{\bibinfo}[2]{#2}
\providecommand{\eprint}[2][]{\url{#2}}

\bibitem{Monroe21}
\bibinfo{author}{Monroe, C.} \emph{et~al.}
\newblock \bibinfo{title}{Programmable quantum simulations of spin systems with
  trapped ions}.
\newblock \emph{\bibinfo{journal}{Rev. Mod. Phys.}}
  \textbf{\bibinfo{volume}{93}}, \bibinfo{pages}{025001}
  (\bibinfo{year}{2021}).
\newblock
  \urlprefix\url{https://link.aps.org/doi/10.1103/RevModPhys.93.025001}.

\bibitem{LukinNature21}
\bibinfo{author}{Ebadi, S.} \emph{et~al.}
\newblock \bibinfo{title}{Quantum phases of matter on a 256-atom programmable
  quantum simulator}.
\newblock \emph{\bibinfo{journal}{Nature}} \textbf{\bibinfo{volume}{595}},
  \bibinfo{pages}{227--232} (\bibinfo{year}{2021}).
\newblock \urlprefix\url{https://doi.org/10.1038/s41586-021-03582-4}.

\bibitem{Pla13}
\bibinfo{author}{Pla, J.~J.} \emph{et~al.}
\newblock \bibinfo{title}{High-fidelity readout and control of a nuclear spin
  qubit in silicon}.
\newblock \emph{\bibinfo{journal}{Nature}} \textbf{\bibinfo{volume}{496}},
  \bibinfo{pages}{334--338} (\bibinfo{year}{2013}).
\newblock \urlprefix\url{https://doi.org/10.1038/nature12011}.

\bibitem{Cory97}
\bibinfo{author}{Cory, D.~G.}, \bibinfo{author}{Fahmy, A.~F.} \&
  \bibinfo{author}{Havel, T.~F.}
\newblock \bibinfo{title}{Ensemble quantum computing by nmr spectroscopy}.
\newblock \emph{\bibinfo{journal}{Proc. Nat. Acad. Sci. (USA)}}
  \textbf{\bibinfo{volume}{94}}, \bibinfo{pages}{1634--1639}
  (\bibinfo{year}{1997}).
\newblock \urlprefix\url{https://www.pnas.org/content/94/5/1634}.

\bibitem{Chuang97}
\bibinfo{author}{Gershenfeld, N.~A.} \& \bibinfo{author}{Chuang, I.~L.}
\newblock \bibinfo{title}{Bulk spin-resonance quantum computation}.
\newblock \emph{\bibinfo{journal}{Science}} \textbf{\bibinfo{volume}{275}},
  \bibinfo{pages}{350--356} (\bibinfo{year}{1997}).
\newblock \urlprefix\url{https://science.sciencemag.org/content/275/5298/350}.

\bibitem{Marcus18}
\bibinfo{author}{Lutchyn, R.~M.} \emph{et~al.}
\newblock \bibinfo{title}{Majorana zero modes in superconductor--semiconductor
  heterostructures}.
\newblock \emph{\bibinfo{journal}{Nature Reviews Materials}}
  \textbf{\bibinfo{volume}{3}}, \bibinfo{pages}{52--68} (\bibinfo{year}{2018}).
\newblock \urlprefix\url{https://doi.org/10.1038/s41578-018-0003-1}.

\bibitem{Arute19}
\bibinfo{author}{Arute, F.} \emph{et~al.}
\newblock \bibinfo{title}{Quantum supremacy using a programmable
  superconducting processor}.
\newblock \emph{\bibinfo{journal}{Nature}} \textbf{\bibinfo{volume}{574}},
  \bibinfo{pages}{505--510} (\bibinfo{year}{2019}).
\newblock \urlprefix\url{https://doi.org/10.1038/s41586-019-1666-5}.

\bibitem{Asaad20}
\bibinfo{author}{Asaad, S.} \emph{et~al.}
\newblock \bibinfo{title}{Coherent electrical control of a single high-spin
  nucleus in silicon}.
\newblock \emph{\bibinfo{journal}{Nature}} \textbf{\bibinfo{volume}{579}},
  \bibinfo{pages}{205--209} (\bibinfo{year}{2020}).
\newblock \urlprefix\url{https://doi.org/10.1038/s41586-020-2057-7}.

\bibitem{Chatterjee21}
\bibinfo{author}{Chatterjee, A.} \emph{et~al.}
\newblock \bibinfo{title}{Semiconductor qubits in practice}.
\newblock \emph{\bibinfo{journal}{Nature Reviews Physics}}
  \textbf{\bibinfo{volume}{3}}, \bibinfo{pages}{157--177}
  (\bibinfo{year}{2021}).
\newblock \urlprefix\url{https://doi.org/10.1038/s42254-021-00283-9}.

\bibitem{Madzik22}
\bibinfo{author}{Madzik, M.~T.} \emph{et~al.}
\newblock \bibinfo{title}{Precision tomography of a three-qubit donor quantum
  processor in silicon}.
\newblock \emph{\bibinfo{journal}{Nature}} \textbf{\bibinfo{volume}{601}},
  \bibinfo{pages}{348--353} (\bibinfo{year}{2022}).
\newblock \urlprefix\url{https://doi.org/10.1038/s41586-021-04292-7}.

\bibitem{Pezzagna21}
\bibinfo{author}{Pezzagna, S.} \& \bibinfo{author}{Meijer, J.}
\newblock \bibinfo{title}{Quantum computer based on color centers in diamond}.
\newblock \emph{\bibinfo{journal}{Applied Physics Reviews}}
  \textbf{\bibinfo{volume}{8}}, \bibinfo{pages}{011308} (\bibinfo{year}{2021}).
\newblock \urlprefix\url{https://doi.org/10.1063/5.0007444}.
\newblock \eprint{https://doi.org/10.1063/5.0007444}.

\bibitem{Rugar21}
\bibinfo{author}{Rugar, A.~E.} \emph{et~al.}
\newblock \bibinfo{title}{Quantum photonic interface for tin-vacancy centers in
  diamond}.
\newblock \emph{\bibinfo{journal}{Phys. Rev. X}} \textbf{\bibinfo{volume}{11}},
  \bibinfo{pages}{031021} (\bibinfo{year}{2021}).
\newblock \urlprefix\url{https://link.aps.org/doi/10.1103/PhysRevX.11.031021}.

\bibitem{nadj2010spin}
\bibinfo{author}{Nadj-Perge, S.}, \bibinfo{author}{Frolov, S.},
  \bibinfo{author}{Bakkers, E.} \& \bibinfo{author}{Kouwenhoven, L.~P.}
\newblock \bibinfo{title}{Spin--orbit qubit in a semiconductor nanowire}.
\newblock \emph{\bibinfo{journal}{Nature}} \textbf{\bibinfo{volume}{468}},
  \bibinfo{pages}{1084--1087} (\bibinfo{year}{2010}).

\bibitem{pla2012single}
\bibinfo{author}{Pla, J.~J.} \emph{et~al.}
\newblock \bibinfo{title}{A single-atom electron spin qubit in silicon}.
\newblock \emph{\bibinfo{journal}{Nature}} \textbf{\bibinfo{volume}{489}},
  \bibinfo{pages}{541--545} (\bibinfo{year}{2012}).

\bibitem{yoneda2018quantum}
\bibinfo{author}{Yoneda, J.} \emph{et~al.}
\newblock \bibinfo{title}{A quantum-dot spin qubit with coherence limited by
  charge noise and fidelity higher than 99.9\%}.
\newblock \emph{\bibinfo{journal}{Nature nanotechnology}}
  \textbf{\bibinfo{volume}{13}}, \bibinfo{pages}{102--106}
  (\bibinfo{year}{2018}).

\bibitem{preskill1998fault}
\bibinfo{author}{Preskill, J.}
\newblock \bibinfo{title}{Fault-tolerant quantum computation}.
\newblock In \emph{\bibinfo{booktitle}{Introduction to quantum computation and
  information}}, \bibinfo{pages}{213--269} (\bibinfo{publisher}{World
  Scientific}, \bibinfo{year}{1998}).

\bibitem{Gottesman18}
\bibinfo{author}{Gottesman, D.}
\newblock \bibinfo{title}{Theory of fault-tolerant quantum computation}.
\newblock \emph{\bibinfo{journal}{Phys. Rev. A}} \textbf{\bibinfo{volume}{57}},
  \bibinfo{pages}{127--137} (\bibinfo{year}{1998}).
\newblock \urlprefix\url{https://link.aps.org/doi/10.1103/PhysRevA.57.127}.

\bibitem{liu2012spin}
\bibinfo{author}{Liu, L.} \emph{et~al.}
\newblock \bibinfo{title}{Spin-torque switching with the giant spin hall effect
  of tantalum}.
\newblock \emph{\bibinfo{journal}{Science}} \textbf{\bibinfo{volume}{336}},
  \bibinfo{pages}{555--558} (\bibinfo{year}{2012}).

\bibitem{pai2012spin}
\bibinfo{author}{Pai, C.-F.} \emph{et~al.}
\newblock \bibinfo{title}{Spin transfer torque devices utilizing the giant spin
  hall effect of tungsten}.
\newblock \emph{\bibinfo{journal}{Applied Physics Letters}}
  \textbf{\bibinfo{volume}{101}}, \bibinfo{pages}{122404}
  (\bibinfo{year}{2012}).

\bibitem{niimi2012giant}
\bibinfo{author}{Niimi, Y.} \emph{et~al.}
\newblock \bibinfo{title}{Giant spin hall effect induced by skew scattering
  from bismuth impurities inside thin film cubi alloys}.
\newblock \emph{\bibinfo{journal}{Physical review letters}}
  \textbf{\bibinfo{volume}{109}}, \bibinfo{pages}{156602}
  (\bibinfo{year}{2012}).

\bibitem{maruyama2009large}
\bibinfo{author}{Maruyama, T.} \emph{et~al.}
\newblock \bibinfo{title}{Large voltage-induced magnetic anisotropy change in a
  few atomic layers of iron}.
\newblock \emph{\bibinfo{journal}{Nature nanotechnology}}
  \textbf{\bibinfo{volume}{4}}, \bibinfo{pages}{158--161}
  (\bibinfo{year}{2009}).

\bibitem{shiota2009voltage}
\bibinfo{author}{Shiota, Y.} \emph{et~al.}
\newblock \bibinfo{title}{Voltage-assisted magnetization switching in ultrathin
  fe80co20 alloy layers}.
\newblock \emph{\bibinfo{journal}{Applied Physics Express}}
  \textbf{\bibinfo{volume}{2}}, \bibinfo{pages}{063001} (\bibinfo{year}{2009}).

\bibitem{shiota2012induction}
\bibinfo{author}{Shiota, Y.} \emph{et~al.}
\newblock \bibinfo{title}{Induction of coherent magnetization switching in a
  few atomic layers of feco using voltage pulses}.
\newblock \emph{\bibinfo{journal}{Nature materials}}
  \textbf{\bibinfo{volume}{11}}, \bibinfo{pages}{39--43}
  (\bibinfo{year}{2012}).

\bibitem{grezes2016ultra}
\bibinfo{author}{Grezes, C.} \emph{et~al.}
\newblock \bibinfo{title}{Ultra-low switching energy and scaling in
  electric-field-controlled nanoscale magnetic tunnel junctions with high
  resistance-area product}.
\newblock \emph{\bibinfo{journal}{Applied Physics Letters}}
  \textbf{\bibinfo{volume}{108}}, \bibinfo{pages}{012403}
  (\bibinfo{year}{2016}).

\bibitem{atulasimha2010bennett}
\bibinfo{author}{Atulasimha, J.} \& \bibinfo{author}{Bandyopadhyay, S.}
\newblock \bibinfo{title}{Bennett clocking of nanomagnetic logic using
  multiferroic single-domain nanomagnets}.
\newblock \emph{\bibinfo{journal}{Applied Physics Letters}}
  \textbf{\bibinfo{volume}{97}}, \bibinfo{pages}{173105}
  (\bibinfo{year}{2010}).

\bibitem{cui2015generation}
\bibinfo{author}{Cui, J.} \emph{et~al.}
\newblock \bibinfo{title}{Generation of localized strain in a thin film
  piezoelectric to control individual magnetoelectric heterostructures}.
\newblock \emph{\bibinfo{journal}{Applied Physics Letters}}
  \textbf{\bibinfo{volume}{107}}, \bibinfo{pages}{092903}
  (\bibinfo{year}{2015}).

\bibitem{d2016experimental}
\bibinfo{author}{D’Souza, N.}, \bibinfo{author}{Salehi~Fashami, M.},
  \bibinfo{author}{Bandyopadhyay, S.} \& \bibinfo{author}{Atulasimha, J.}
\newblock \bibinfo{title}{Experimental clocking of nanomagnets with strain for
  ultralow power boolean logic}.
\newblock \emph{\bibinfo{journal}{Nano letters}} \textbf{\bibinfo{volume}{16}},
  \bibinfo{pages}{1069--1075} (\bibinfo{year}{2016}).

\bibitem{mathurin2016stress}
\bibinfo{author}{Mathurin, T.} \emph{et~al.}
\newblock \bibinfo{title}{Stress-mediated magnetoelectric control of
  ferromagnetic domain wall position in multiferroic heterostructures}.
\newblock \emph{\bibinfo{journal}{Applied Physics Letters}}
  \textbf{\bibinfo{volume}{108}}, \bibinfo{pages}{082401}
  (\bibinfo{year}{2016}).

\bibitem{heron2011electric}
\bibinfo{author}{Heron, J.} \emph{et~al.}
\newblock \bibinfo{title}{Electric-field-induced magnetization reversal in a
  ferromagnet-multiferroic heterostructure}.
\newblock \emph{\bibinfo{journal}{Physical review letters}}
  \textbf{\bibinfo{volume}{107}}, \bibinfo{pages}{217202}
  (\bibinfo{year}{2011}).

\bibitem{d2018energy}
\bibinfo{author}{D’Souza, N.} \emph{et~al.}
\newblock \bibinfo{title}{Energy-efficient switching of nanomagnets for
  computing: straintronics and other methodologies}.
\newblock \emph{\bibinfo{journal}{Nanotechnology}}
  \textbf{\bibinfo{volume}{29}}, \bibinfo{pages}{442001}
  (\bibinfo{year}{2018}).

\bibitem{wang2015magnetoelectric}
\bibinfo{author}{Wang, K.~L.}, \bibinfo{author}{Lee, H.} \&
  \bibinfo{author}{Amiri, P.~K.}
\newblock \bibinfo{title}{Magnetoelectric random access memory-based circuit
  design by using voltage-controlled magnetic anisotropy in magnetic tunnel
  junctions}.
\newblock \emph{\bibinfo{journal}{IEEE Transactions on Nanotechnology}}
  \textbf{\bibinfo{volume}{14}}, \bibinfo{pages}{992--997}
  (\bibinfo{year}{2015}).

\bibitem{nowak2016dependence}
\bibinfo{author}{Nowak, J.~J.} \emph{et~al.}
\newblock \bibinfo{title}{Dependence of voltage and size on write error rates
  in spin-transfer torque magnetic random-access memory}.
\newblock \emph{\bibinfo{journal}{IEEE Magnetics Letters}}
  \textbf{\bibinfo{volume}{7}}, \bibinfo{pages}{1--4} (\bibinfo{year}{2016}).

\bibitem{labanowski2018voltage}
\bibinfo{author}{Labanowski, D.} \emph{et~al.}
\newblock \bibinfo{title}{Voltage-driven, local, and efficient excitation of
  nitrogen-vacancy centers in diamond}.
\newblock \emph{\bibinfo{journal}{Science advances}}
  \textbf{\bibinfo{volume}{4}}, \bibinfo{pages}{eaat6574}
  (\bibinfo{year}{2018}).

\bibitem{wang2020electrical}
\bibinfo{author}{Wang, X.} \emph{et~al.}
\newblock \bibinfo{title}{Electrical control of coherent spin rotation of a
  single-spin qubit}.
\newblock \emph{\bibinfo{journal}{npj Quantum Information}}
  \textbf{\bibinfo{volume}{6}}, \bibinfo{pages}{1--6} (\bibinfo{year}{2020}).

\bibitem{meso1}
\bibinfo{author}{Mirkamali, M.~S.} \& \bibinfo{author}{Cory, D.~G.}
\newblock \bibinfo{title}{Mesoscopic spin systems as quantum entanglers}.
\newblock \emph{\bibinfo{journal}{Phys. Rev. A}}
  \textbf{\bibinfo{volume}{101}}, \bibinfo{pages}{032320}
  (\bibinfo{year}{2020}).
\newblock \urlprefix\url{https://link.aps.org/doi/10.1103/PhysRevA.101.032320}.

\bibitem{meso2}
\bibinfo{author}{Zhukov, A.~A.}, \bibinfo{author}{Shapiro, D.~S.},
  \bibinfo{author}{Pogosov, W.~V.} \& \bibinfo{author}{Lozovik, Y.~E.}
\newblock \bibinfo{title}{Dynamics of a mesoscopic qubit ensemble coupled to a
  cavity: Role of collective dark states}.
\newblock \emph{\bibinfo{journal}{Phys. Rev. A}} \textbf{\bibinfo{volume}{96}},
  \bibinfo{pages}{033804} (\bibinfo{year}{2017}).
\newblock \urlprefix\url{https://link.aps.org/doi/10.1103/PhysRevA.96.033804}.

\bibitem{meso3}
\bibinfo{author}{Barbara, B.}
\newblock \bibinfo{title}{Mesoscopic systems: classical irreversibility and
  quantum coherence}.
\newblock \emph{\bibinfo{journal}{Philosophical Transactions of the Royal
  Society A: Mathematical, Physical and Engineering Sciences}}
  \textbf{\bibinfo{volume}{370}}, \bibinfo{pages}{4487--4516}
  (\bibinfo{year}{2012}).

\bibitem{meso4}
\bibinfo{author}{Gangloff, D.~A.} \emph{et~al.}
\newblock \bibinfo{title}{Witnessing quantum correlations in a nuclear ensemble
  via an electron spin qubit}.
\newblock \emph{\bibinfo{journal}{Nature Physics}}
  \textbf{\bibinfo{volume}{17}}, \bibinfo{pages}{1247--1253}
  (\bibinfo{year}{2021}).
\newblock \urlprefix\url{https://doi.org/10.1038/s41567-021-01344-7}.

\bibitem{meso5}
\bibinfo{author}{Giedke, G.}, \bibinfo{author}{Taylor, J.~M.},
  \bibinfo{author}{D'Alessandro, D.}, \bibinfo{author}{Lukin, M.~D.} \&
  \bibinfo{author}{Imamo\ifmmode~\breve{g}\else \u{g}\fi{}lu, A.}
\newblock \bibinfo{title}{Quantum measurement of a mesoscopic spin ensemble}.
\newblock \emph{\bibinfo{journal}{Phys. Rev. A}} \textbf{\bibinfo{volume}{74}},
  \bibinfo{pages}{032316} (\bibinfo{year}{2006}).
\newblock \urlprefix\url{https://link.aps.org/doi/10.1103/PhysRevA.74.032316}.

\bibitem{meso6}
\bibinfo{author}{Beterov, I.~I.} \emph{et~al.}
\newblock \bibinfo{title}{Coherent control of mesoscopic atomic ensembles for
  quantum information}.
\newblock \emph{\bibinfo{journal}{Laser Physics}}
  \textbf{\bibinfo{volume}{24}}, \bibinfo{pages}{074013}
  (\bibinfo{year}{2014}).
\newblock \urlprefix\url{https://doi.org/10.1088/1054-660x/24/7/074013}.

\bibitem{meso7}
\bibinfo{author}{Gangloff, D.~A.} \emph{et~al.}
\newblock \bibinfo{title}{Witnessing quantum correlations in a nuclear spin
  ensemble via a proxy qubit}.
\newblock In \emph{\bibinfo{booktitle}{Quantum Information and Measurement VI
  2021}}, \bibinfo{pages}{Tu3A.2} (\bibinfo{publisher}{Optica Publishing
  Group}, \bibinfo{year}{2021}).

\bibitem{meso8}
\bibinfo{author}{Jackson, D.~M.} \emph{et~al.}
\newblock \bibinfo{title}{Quantum sensing of a coherent single spin excitation
  in a nuclear ensemble}.
\newblock \emph{\bibinfo{journal}{Nature Physics}} \bibinfo{pages}{585--590}.

\bibitem{meso9}
\bibinfo{author}{Rabl, P.} \emph{et~al.}
\newblock \bibinfo{title}{Hybrid quantum processors: Molecular ensembles as
  quantum memory for solid state circuits}.
\newblock \emph{\bibinfo{journal}{Phys. Rev. Lett.}}
  \textbf{\bibinfo{volume}{97}}, \bibinfo{pages}{033003}
  (\bibinfo{year}{2006}).

\bibitem{Bhattacharya18}
\bibinfo{author}{Bhattacharya, D.} \& \bibinfo{author}{Atulasimha, J.}
\newblock \bibinfo{title}{Skyrmion-mediated voltage-controlled switching of
  ferromagnets for reliable and energy-efficient two-terminal memory}.
\newblock \emph{\bibinfo{journal}{ACS Applied Materials \& Interfaces}}
  \textbf{\bibinfo{volume}{10}}, \bibinfo{pages}{17455--17462}
  (\bibinfo{year}{2018}).
\newblock \urlprefix\url{https://doi.org/10.1021/acsami.8b02791}.

\bibitem{Knill05}
\bibinfo{author}{Knill, E.}
\newblock \bibinfo{title}{Quantum computing with realistically noisy devices}.
\newblock \emph{\bibinfo{journal}{Nature}} \textbf{\bibinfo{volume}{434}},
  \bibinfo{pages}{39--44} (\bibinfo{year}{2005}).
\newblock \urlprefix\url{https://doi.org/10.1038/nature03350}.

\bibitem{Veldhorst14}
\bibinfo{author}{Veldhorst, M.} \emph{et~al.}
\newblock \bibinfo{title}{An addressable quantum dot qubit with fault-tolerant
  control-fidelity}.
\newblock \emph{\bibinfo{journal}{Nature Nanotechnology}}
  \textbf{\bibinfo{volume}{9}}, \bibinfo{pages}{981--985}
  (\bibinfo{year}{2014}).
\newblock \urlprefix\url{https://doi.org/10.1038/nnano.2014.216}.

\bibitem{Morton16}
\bibinfo{author}{Wolfowicz, G.} \& \bibinfo{author}{Morton, J.}
\newblock \bibinfo{title}{Pulse techniques for quantum information processing}.
\newblock \emph{\bibinfo{journal}{eMagRes}} \textbf{\bibinfo{volume}{5}},
  \bibinfo{pages}{1515--1528} (\bibinfo{year}{2016}).

\bibitem{gottesman1997stabilizer}
\bibinfo{author}{Gottesman, D.}
\newblock \emph{\bibinfo{title}{Stabilizer codes and quantum error correction}}
  (\bibinfo{publisher}{California Institute of Technology},
  \bibinfo{year}{1997}).

\bibitem{BOWDREY02}
\bibinfo{author}{Bowdrey, M.~D.}, \bibinfo{author}{Oi, D.~K.},
  \bibinfo{author}{Short, A.~J.}, \bibinfo{author}{Banaszek, K.} \&
  \bibinfo{author}{Jones, J.~A.}
\newblock \bibinfo{title}{Fidelity of single qubit maps}.
\newblock \emph{\bibinfo{journal}{Physics Letters A}}
  \textbf{\bibinfo{volume}{294}}, \bibinfo{pages}{258--260}
  (\bibinfo{year}{2002}).
\newblock
  \urlprefix\url{https://www.sciencedirect.com/science/article/pii/S0375960102000695}.

\bibitem{LEVITT86}
\bibinfo{author}{Levitt, M.~H.}
\newblock \bibinfo{title}{Composite pulses}.
\newblock \emph{\bibinfo{journal}{Progress in Nuclear Magnetic Resonance
  Spectroscopy}} \textbf{\bibinfo{volume}{18}}, \bibinfo{pages}{61--122}
  (\bibinfo{year}{1986}).
\newblock
  \urlprefix\url{https://www.sciencedirect.com/science/article/pii/007965658680005X}.

\bibitem{Ryan10}
\bibinfo{author}{Ryan, C.~A.}, \bibinfo{author}{Hodges, J.~S.} \&
  \bibinfo{author}{Cory, D.~G.}
\newblock \bibinfo{title}{Robust decoupling techniques to extend quantum
  coherence in diamond}.
\newblock \emph{\bibinfo{journal}{Phys. Rev. Lett.}}
  \textbf{\bibinfo{volume}{105}}, \bibinfo{pages}{200402}
  (\bibinfo{year}{2010}).
\newblock
  \urlprefix\url{https://link.aps.org/doi/10.1103/PhysRevLett.105.200402}.

\bibitem{Vansteenkiste14}
\bibinfo{author}{Vansteenkiste, A.} \emph{et~al.}
\newblock \bibinfo{title}{The design and verification of mumax3}.
\newblock \emph{\bibinfo{journal}{AIP Advances}} \textbf{\bibinfo{volume}{4}},
  \bibinfo{pages}{107133} (\bibinfo{year}{2014}).
\newblock \urlprefix\url{https://doi.org/10.1063/1.4899186}.
\newblock \eprint{https://doi.org/10.1063/1.4899186}.

\bibitem{Yang11}
\bibinfo{author}{Yang, H.~X.} \emph{et~al.}
\newblock \bibinfo{title}{First-principles investigation of the very large
  perpendicular magnetic anisotropy at fe$|$mgo and co$|$mgo interfaces}.
\newblock \emph{\bibinfo{journal}{Phys. Rev. B}} \textbf{\bibinfo{volume}{84}},
  \bibinfo{pages}{054401} (\bibinfo{year}{2011}).
\newblock \urlprefix\url{https://link.aps.org/doi/10.1103/PhysRevB.84.054401}.

\bibitem{Niranjan10}
\bibinfo{author}{Niranjan, M.~K.}, \bibinfo{author}{Duan, C.-G.},
  \bibinfo{author}{Jaswal, S.~S.} \& \bibinfo{author}{Tsymbal, E.~Y.}
\newblock \bibinfo{title}{Electric field effect on magnetization at the
  fe/mgo(001) interface}.
\newblock \emph{\bibinfo{journal}{Applied Physics Letters}}
  \textbf{\bibinfo{volume}{96}}, \bibinfo{pages}{222504}
  (\bibinfo{year}{2010}).
\newblock \urlprefix\url{https://doi.org/10.1063/1.3443658}.
\newblock \eprint{https://doi.org/10.1063/1.3443658}.

\bibitem{Amiri12}
\bibinfo{author}{AMIRI, P.~K.} \& \bibinfo{author}{WANG, K.~L.}
\newblock \bibinfo{title}{Voltage-controlled magnetic anisotropy in spintronic
  devices}.
\newblock \emph{\bibinfo{journal}{SPIN}} \textbf{\bibinfo{volume}{02}},
  \bibinfo{pages}{1240002} (\bibinfo{year}{2012}).
\newblock \urlprefix\url{https://doi.org/10.1142/S2010324712400024}.
\newblock \eprint{https://doi.org/10.1142/S2010324712400024}.

\bibitem{Wang12}
\bibinfo{title}{Electric-field-assisted switching in magnetic tunnel
  junctions}.
\newblock \emph{\bibinfo{journal}{Nature Materials}}
  \textbf{\bibinfo{volume}{11}}, \bibinfo{pages}{64--68}
  (\bibinfo{year}{2012}).
\newblock \urlprefix\url{https://doi.org/10.1038/nmat3171}.

\bibitem{li_lee_razavi_18}
\bibinfo{author}{Li, X.}, \bibinfo{author}{Lee, A.}, \bibinfo{author}{Razavi,
  S.~A.}, \bibinfo{author}{Wu, H.} \& \bibinfo{author}{Wang, K.~L.}
\newblock \bibinfo{title}{Voltage-controlled magnetoelectric memory and logic
  devices}.
\newblock \emph{\bibinfo{journal}{MRS Bulletin}} \textbf{\bibinfo{volume}{43}},
  \bibinfo{pages}{970--977} (\bibinfo{year}{2018}).

\end{thebibliography}

\section{Acknowledgements}
JA, MFC and MMR were supported in part by National Science Foundation (NSF) grants 1815033 and 1909030.  The research at UCLA was partially supported by NSF awards CHE-2002313 and 1936375.

\section{Competing interests}
The authors declare no competing interests.


\section{Author contributions}
JA, KLW, RNS and LSB conceived the idea. All authors discussed the results and commented on the paper. JA defined the nanomagnet magnetization control and field inhomogeneity problem and MFC performed the micromagnetic simulations with help from MMR and WAM. LSB defined the spin evolution problem and MN performed the spin dynamics simulations and quantum gate calculations. JA, MFC, LSB and MN wrote the paper.     $^\dagger$These authors (MN, MFC) contributed equally.
\end{document}